# A Chaotic Cipher Mmohocc and Its Security Analysis

**Xiaowen Zhang[1], Li Shu[2], Ke Tang[1]**

[1] Dept. of CS, the Graduate Center, CUNY
[2] College of Computer Science, Sichuan University
xzhang2(at)gc.cuny.edu

**Abstract**
In this paper we introduce a new chaotic stream cipher Mmohocc which utilizes the fundamental chaos characteristics. The designs of the major components of the cipher are given. Its cryptographic properties of period, auto- and cross-correlations, and the mixture of Markov processes and spatiotemporal effects are investigated. The cipher is resistant to the related-key-IV attacks, Time/Memory/Data tradeoff attacks, algebraic attacks, and chosen-text attacks. The keystreams successfully passed two batteries of statistical tests and the encryption speed is comparable with RC4.

## 1   Introduction

Claude E. Shannon [Shannon 1949] indicated that stretch-and-fold mechanism of mixing transformation is an important component of a cipher and it is also a way to confuse and diffuse messages. The chaotic dynamical systems, which exhibit unpredictable, mixing, ergodic, and extremely sensitive to initial conditions, have shown strong connection with cryptography since its inception. It has attracted a broad interest among cryptographic researchers to investigate the cryptographic applications for chaotic properties [Baptista 1998, Alvarez 1999, Jakimoski 2001, Fridrich 1998, Kocarev 1998, Argyris1 2005]. However the properties of normal chaos systems are based on continuous systems and they are long-term behaviors. Due to limited resources and computational power in digital systems those properties could be severely degraded or even disappeared. In order to use chaotic properties in cryptosystems, we have devised a technique which accelerated chaotic behaviors more rapidly than those in a normal chaos system.

The Mmohocc [Zhang 2006] (pronounced "mow-hock"), an acronym for *m*ulti-*m*ap *o*rbit *ho*pping *c*haotic *c*ipher, is still under development. It is a software stream cipher without authentication mechanism and satisfies the Profile-1 criteria for the ECRYPT Stream Cipher Project [Ecrypt 2005]. Its primary goal is to exploit basic properties of chaos systems to design a cryptographically strong, fast, and feasible cipher. By using multiple maps, i.e., multiple chaos systems, the Mmohocc produces an extremely long chaotic sequence inspired by Vernam's one-time pad. By hopping between multiple orbits generated from multiple maps the cipher obtains its confusion and diffusion properties in a much faster way, it is a basic asset for any good cryptosystem.

For any stream cipher at each time unit, e.g. a clock tick for Linear Feedback Shift Register (LFSR)-based stream cipher, the keystream unit produced by the cipher is obtained by applying a filtering/output function to the current internal state. The internal state is then updated by a state transition function. Both filtering function and transition function must be chosen carefully in order to make the underlying cipher resistant to known attacks [Canteaut 2005]. In our Mmohocc cipher the filtering function is an S-Box-like function and the state transition function is a bank of multiple chaotic maps.

A chaotic map $F$ is a map/function, usually a non-linear discrete dynamical iteration equation, which exhibits some sort of chaotic behavior. A chaotic orbit is the trajectory that a chaotic map iterated. Given $x_0 \in R$ and a chaotic map $F$, we define the orbit of $x_0$ under $F$ to be the sequence of points $x_0$, $x_1 = F(x_0)$, $x_2 = F^2(x_0)$, $x_n = F^n(x_0)$, …. The point $x_0$ is called the seed of the orbit [Devaney 1992].

The rest of the paper is organized as follows. Section 2 we describe the construction, major components, and pseudo code implementation of the Mmohocc cipher. In Section 3 the cryptographic properties and security analysis against some attacks are investigated. Section 4 we briefly present randomness test results and encryption speed performance against RC4. Finally the conclusions are drawn in Section 5.

## 2   The Mmohocc – A Chaotic Stream Cipher

The Mmohocc "leaps" among substantially many chaotic orbits to "pick up" its random sequence points. The base ground for random sequences is spread among lots of chaotic orbits, which are generated from multiple chaotic maps. The notion of orbit-hopping is also inspired from the frequency-hopping (FH) communication. It is one of the modulation methods in spread spectrum communications, in which radio signals, following a hopping pattern, switch a carrier among many frequencies. A hopping pattern is a predefined pseudorandom sequence known to both transmitter and receiver [Wikipedia].

Fig. 1 shows the construction blocks of the Mmohocc. The Key Scheduling expands a 128-bit (or 256-bit, or 512-bit) key into 8 (or 16) subkeys (SKs) for controlling the operations of multiple chaotic maps. Each subkey contains certain number of fields. Among them a field called hopping-pattern serial number (hpsn) is used by the Hopping Mechanism to govern the jumping behavior among the chaotic orbits. In the following subsections we will explain some of the major building components of the cipher.

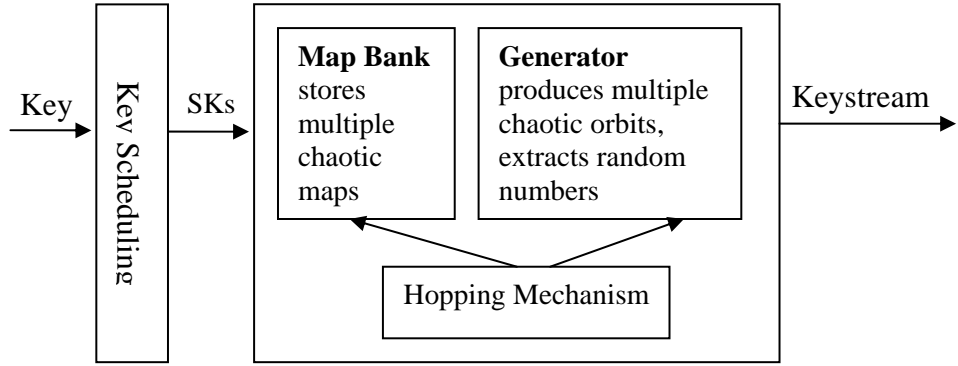

Fig. 1  Block Diagram of the Mmohocc Cipher

## 2.1  Chaotic Map Bank

For the pseudo random number generator of the stream cipher, we are using a set of chaotic maps from the Map Bank whose parameters are properly tuned to ensure that all maps lead to chaotic. A map is chaotic [Wikipedia] if it is sensitive to initial conditions, topologically mixing, and its periodic orbits are dense.

In our experimental system, we are using a variety of logistic maps and quadratic maps. Both types have been well studied and understood [Devaney 1992]. The logistic map is defined as

$$x_{n+1} = rx_n(1-x_n), \; x_n \in (0,\ 1)$$

where $0 \leq r \leq 4$. Most values of $r$ beyond 3.57 exhibit chaotic behavior (but there are still certain isolated values of $r$ that appear to show non-chaotic behavior, e.g. $r = 3.82$).

The quadratic map is defined as $x_{n+1} = x_n^2 + c$ and its behavior depends on parameter $c$. Most values of $c$ beyond –1.45 left exhibit chaotic behaviors. When $c$ is close to –2 the orbits $x_n$ distribute within (–2, +2), i.e. $x_n \in (-2,\ +2)$. We choose $c$ between –1.9 to –2 for our set of quadratic maps.

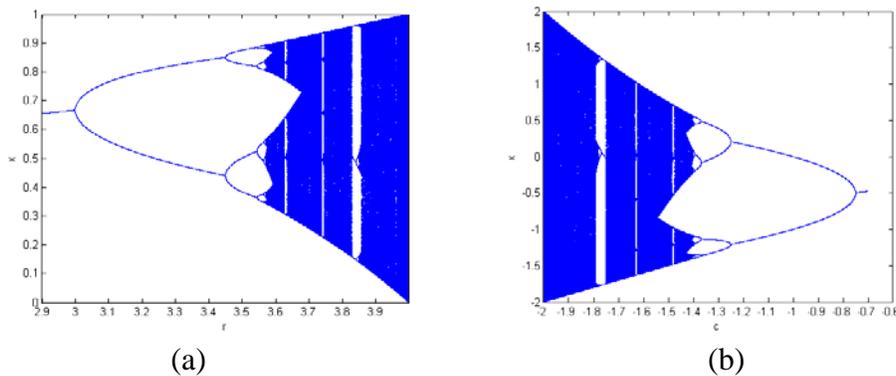

(a)          (b)

Fig. 2  Logistic Maps (a) and Quadratic Maps (b)
(Using the same initial value $x_0 = 0.1$ and $x_0 = 0$ for all $r$'s and all $c$'s, respectively.)

In the implementation, for logistic and quadratic maps we choose $r$ and $c$ properly to avoid windows (openings), see Fig. 2.

### 2.2  Key, Key Scheduling, and Initialization Vector

Currently Mmohocc has three versions, each with 128-, 256-, and 512-bit key, we call them Mmohocc-128, Mmohocc-256, and Mmohocc-512, respectively.

The Mmohocc-128 / 256 uses 8 chaotic maps (#maps = 8) and a 64 bits initialization vector (IV). This is like it has 8 rounds in parallel (simultaneously). Each round needs a subkey to control the behavior of the corresponding map. Key scheduling is as follows.

Step-1: Halve the key and IV. The key is written in the form $K_L K_R$, where $K_L$ consists of the first 64 / 128 bits and $K_R$ consists of the last 64 / 128 bits. And IV is written in the form $V_L V_R$, where $V_L$ consists of the first 32 bits and $V_R$ consists of the last 32 bits.

Step-2: Concatenate the halves as $V_L K_L$ and $V_R K_R$. They are now 96 / 160 bits each.

Step-3: Expand the two concatenations. Hash $V_L K_L$ and $V_R K_R$ with $SHA$256, we get two 256-bit hashes $H_L = SHA256(V_L K_L)$ and $H_R = SHA256(V_R K_R)$.

Step-4: Interweave and split. Interweave $H_L$ and $H_R$ bit-by-bit to get a 512-bit long string, then split it into 8 even substrings each with 64-bit $k_1$, $k_2$, …, $k_8$. They are 8 subkeys for 8 chaotic maps, correspondingly.

For Mmohocc-512, the key scheduling is similar to Mmohocc-128/256, the difference is that it uses 16 chaotic maps and a 128-bit IV.

This key scheduling provides a fine chopping to the input key. Sufficiently use the key and IV information. We expand the key into subkeys, does not reduce the key space.

For our system, IVs are multiple 64 bits. For Mmohocc-128 / 256 the IV is 64-bit; for Mmohocc-512, the IV is 128-bit.

### 2.3  Hopping Patterns

The hopping pattern is a predefined pseudorandom sequence of certain number of orbits, it tells how Mmohocc hops among orbits. In the current version the hpsn field of a subkey takes 8 bits long, therefore it can be changed between 0 and 255. Each hpsn corresponds to one of available hopping patterns, which are stored in a lookup table. For instance, if a chaotic map has 11 orbits and its hpsn is 141, then the hopping-pattern is {2, 1, 9, 11, 10, 6, 5, 8, 7, 4, 3} is returned from the lookup table. We know that when it comes to this map Mmohocc extracts random numbers on orbit 2 first and then orbit 1, orbit 9, … orbit 3 in this order. For more details, see Table 2 in Appendix A.

### 2.4 Random Number Extraction

A chaotic orbit point $x_n$ is a real number (in the case of logistic map, $x_n < 1.0$) and is represented in a data type of double in computer. We extract two smaller integers as random numbers from the point $x_n$ in the following method, see Fig. 3. By doing so, we further muddle the bits and add additional randomness to the keystream. This is the filtering function of the cipher, a multi-output Boolean function. It acts like an S-Box with 32-bit-in and 16-bit-out.

- Step 1: Convert double type $x_n$ into a 32-bit integer by multiplying $2^{52}$.
- Step 2: Split the above integer into four 8-bit integers $a$, $b$, $c$, $d$. Here $a$ and $d$ are the most left and right 8-bit, $b$ and $c$ are in the middle. Then the output two smaller pseudorandom integers are: $p = a \oplus c$ and $q = b \oplus d$.

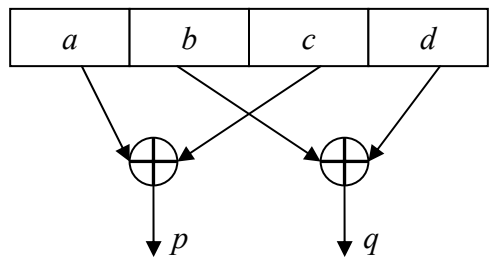

Fig. 3   Random Number Extraction

### 2.5 Pseudo-code Implementation

Here is the pseudo-code implementation for the main part of the Mmohocc cipher.

```
keyScheduling()
initializeMaps(seeds[i],offsets[i])
initializeHoppingPatternArray(hopp[i][j])
for_1 i=0 to NUM_OF_MAPS    //maps
   for_2 j=0 to NUM_OF_ORBITS_OF_MAP_i   //orbits
      startOrbit = seeds[i]+hopp[i][j]*offsets[i]*hopping_skew;
      for_3 s to SETTLES_OF_MAP_i      // settles
         //iterate MAP_i SETTLES_OF_MAP_i times
      end_for_3
      for_4 k=0 to SAMPLES_OF_MAP_i    // samples
         //iterate MAP_i once
         generateKeystream(keystream)
         ciphertext = keystream + plaintext
      end_for_4
   end_for_2
end_for_1
```

## 3   Security Analysis to the Mmohocc Cipher

In this section we investigate the period, auto- and cross-correlations of keystreams, describe the cipher in Markov process and spatiotemporal lattices, and cryptanalyze the security of the cipher against some well known attacks on stream ciphers.

### 3.1   Period

We all know that any chaotic orbit will eventually become periodic in computer realization with a finite precision. However, when we have many chaotic maps and each map generates sufficiently large number of chaotic orbits, the period of the mixture of the orbits will be extremely large.

State space $S$ in bits is (#maps * #orbits * 32). If #maps = 16, #orbits = 20 ~ 35. In average the state space is 16*27*32 = 13824-bit. This ensures that the period of the keystream is extremely big. But the exact period of Mmohocc is difficult to predict, it also depends on selection of map coefficients and other coefficients in the implementation, for example, the "hopping skew" to offset[$i$] (0.19, 0.29, or other small numbers bigger than 0.01). The average period of the keystream is approximately $2^{13824}$.

### 3.2   Auto- and Cross-correlations

The cross-correlation is a measure of similarity of two bit sequences. It is a function of the relative time/lag between the sequences. Large cross-correlations between sequences mean that the sequences are very similar to each other and usually not statistically independent. The auto-correlation of a sequence $s_n$ gives the amount of similarity between the sequence $s_n$ and a shift of $s_n$ by $m$ positions. It is simply the cross-correlation of the sequence against a time-shifted version of itself. In general it is desirable to have as small values as possible for the correlation (except for $m = 0$ in the auto-correlation) to ensure independence of sequences. It is that the auto- and cross-correlations are $\delta$-like function and close-to-zero, respectively.

If we treat the keystreams produced by the Mmohocc cipher as bit sequences, we can examine the similarities between the sequences by calculating the auto- and cross-correlations. We take bit sequence length $N$, sliding from $-N/2$ to $+N/2$.

We use the following correlation definition formulas to calculate the auto- and cross-correlations. The auto-correlation $AC$ at lag $m$ for the sequence $s_n(i)$ is (here $n+m$ is performed modulo $N$)

$$AC_i(m) = \lim_{N \to \infty} \frac{1}{N} \sum_{n=1}^{N} [s_n(i) - \mu(i)][s_{n+m}(i) - \mu(i)].$$

And the cross-correlation $CC$ at lag $m$ for the sequences $s_n(i)$ and $s_n(j)$ is

$$CC_{ij}(m) = \lim_{N \to \infty} \frac{1}{N} \sum_{n=1}^{N} [s_n(i) - \mu(i)][s_{n+m}(j) - \mu(j)],$$

where $\mu(i)$ and $\mu(j)$ are the averages (expected values) for $s_n(i)$ and $s_n(j)$, respectively. Here are two examples from the experiment and we find that the auto-correlation is and.

Fig. 4 is obtained by arbitrarily choosing two bit sequence segments (the top two for $N=2048$ bits and the bottom two for $N=8192$ bits) from two long different keystreams. The average auto-correlation $AC(m)$ is very close to 0 everywhere except at $m=0$, i.e. a $\delta$-like function. The cross-correlation $CC(m)$ is close-to-zero everywhere. Also we can see that the longer the sequences, the less correlations are between the sequences.

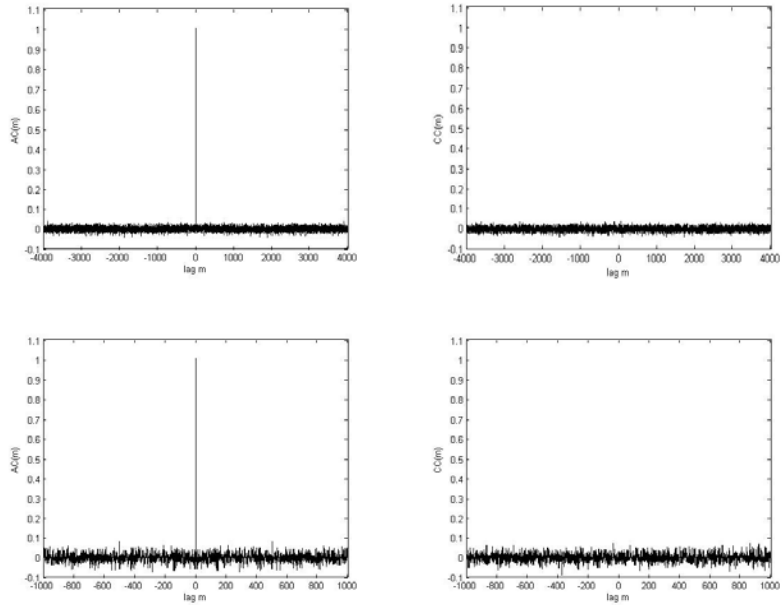

Fig. 4  Auto- and Cross-Correlation for $N=8192$ and $N=2048$

### 3.3  Mixture of Markov Processes and Spatiotemporal Effects

The cipher system is deterministic as long as the nonlinear equations are defined, however, it is a rather complicated mixture of multiple Markov processes. In the Mmohocc-128 cipher there are 8 chaotic maps

$$s_{i,n+1} = F_0(s_{i,n}), \; t_{i,n+1} = F_1(t_{i,n}), \; u_{i,n+1} = F_2(u_{i,n}), \; v_{i,n+1} = F_3(v_{i,n}),$$
$$w_{i,n+1} = F_4(w_{i,n}), \; x_{i,n+1} = F_5(x_{i,n}), \; y_{i,n+1} = F_6(y_{i,n}), \; z_{i,n+1} = F_7(z_{i,n}),$$

where $F0 \sim F7$ are 8 chaotic maps, $n$ is time index and $i$ is orbit index. The sequence of iterative equations produces a stochastic process, a mixture of 8 Markov processes due to hopping and mixing process in the finite state space. This makes the cycle of the mixed Markov process even longer [Freedman 1962].

For instance, when the system iterates to chaotic map $F_1$, according to the subkey for the map #orbits = 11, #samples = 17, and its hopping-pattern is {7, 3, 9, 1, 6, 10, 4, 5, 2, 11, 8}. Then when it comes to this map the Mmohocc picks 17 points on orbit 7 first ($t_{7,0}$, $t_{7,1}$, …, $t_{7,16}$) and then another 17 points on orbit 3 ($t_{3,0}$, $t_{3,1}$, …, $t_{3,16}$), then on orbit 9 ($t_{9,0}$, $t_{9,1}$, …, $t_{9,16}$), … , on orbit 11 ($t_{11,0}$, $t_{11,1}$, …, $t_{11,16}$) in this order. If we put them in a 2-D array, it will be a *#orbits* rows and *#samples* columns points array

$$\begin{pmatrix} t_{7,0}, & t_{7,1}, & \cdots, & t_{7,16} \\ t_{3,0}, & t_{3,1}, & \cdots, & t_{3,16} \\ \vdots & \vdots & \ddots & \vdots \\ t_{8,0}, & t_{8,1}, & \cdots, & t_{8,16} \end{pmatrix}.$$

Or we can represent it in a 1-D array as

$$(t_{7,0}, t_{7,1}, \cdots, t_{7,16}, t_{3,0}, t_{3,1}, \cdots, t_{3,16}, \cdots, t_{8,0}, t_{8,1}, \cdots t_{8,16}).$$

Then according to the aforementioned random number extraction method, the Mmohocc generates (*#orbits* x *#samples*) random numbers. They constitute a segment of the random numbers. All segments constitute a much complicated keystream with extremely large period. On different map, *#orbits*, *#samples*, and the corresponding hopping-pattern are different. The three factors determine the random number extraction base array.

If we treat each chaotic orbit as one dimension in space, then our Mmohocc cipher will be a spatiotemporal dynamical system. It exhibits chaotic properties in both space and time.

Multiple maps and multiple orbits spread chaotic properties to multi-dimensional directions. This property makes the Mmohocc cipher a collection of multiple spatiotemporal chaotic systems without coupling. In general, a spatiotemporal chaotic system is described by CML (Coupled Map Lattice) model [Willeboordse 1994].

$$x_{n+1}(i) = (1-\varepsilon)f(x_n(i)) + \varepsilon f(x_n(i-1))$$

where *n* is the time index, $i = 1, 2, …, N$ is the space/lattice index (*N* is the lattice size), and $\varepsilon$ the coupling coefficient. The *f* is mapping function, i.e., the chaotic map. For $\varepsilon \to 0$, i.e.

$$x_{n+1}(i) = f(x_n(i))$$

there is no coupling. Therefore local neighborhoods have no influence on the behavior of the CML. This situation represents independently operating local orbits at each lattice site. In the multi-map orbit hopping situation there are multiple mapping functions involved $F_0$, $F_1$, …, $F_k$. And the size of the lattice $N$ (number of orbits) is a variable, it depends on the key.

Besides the speed advantage by parallel computation, the spatiotemporal chaotic system further enhances time and space mixing property.

### 3.4   Security Against Attacks

An IV allows a stream cipher to produce a unique stream independent from other streams produced by the same encryption key, without having to go through a re-keying process [Wikipedia]. In the key scheduling procedure the hash function guarantees that the key plus IV is chopped into subkeys in a very complicated manner. Even a simple incremental IV is used, the cipher is still resistant to related-IV attacks. This cipher is also immune to the subkey guessing and related-key attacks, because the key is not simply applied to chaotic maps instead it is expanded into subkeys by secure hash functions. Obtaining a few subkeys has no help on deducing the key unless the key scheduling procedure is totally insecure. Although this costs a little bit more time, this is just a one-time initialization procedure.

The Mmohocc cipher is resistant to Time/Memory/Data tradeoff attack. Such an attack has two phases: in the preprocessing phase an adversary explores the structure of the stream cipher and summarizes his findings in large tables. In the real attack phase the adversary uses the pre-computed tables and actual data produced from a particular unknown key to find the secret key quickly [Biryukov 2000]. The TMD complexity of the cipher is much larger than $2^{128}$. The large number of states (period $2^{13824}$ is large enough) eliminates the threat of the attack. Also from [Babbage 1996] if the state space is much larger than the number of secret keys, then this attack will provide no improvement. The extended Time/Memory/Key tradeoff attack [Biryukov 2005] can be prevented by using keys longer than 128-bit.

Algebraic attacks on stream ciphers [Courtois 2003, 2003, 2005] recover the key by solving an overdefined system of multivariate equations. Such attacks can break LFSR-based stream ciphers with linear feedback when the output is obtained by a Boolean function. The Mmohocc is a chaotic cipher and the transition function has strong non-linear properties and it is immune to this kind of attacks by nature. Besides, the cipher is using an S-Box-like filtering function with 32-bit input and 16-bit output.

There are other attacks to be considered here. The ciphertext-only attack tries to use the output cipher to recover the key; it is in some sense equivalent to brute force attack. The key space for the cipher is large enough, at least $2^{128}$. So it is not feasible to apply these attacks to the cipher.

The known-plaintext, chosen-plaintext, and chosen-ciphertext attacks will be carefully checked here. A known-plaintext attack is one that an adversary knows one or more plaintexts and the corresponding ciphertexts. Chosen-plaintext/ciphertext attack implies that an adversary has access to the encryption/decryption equipment, therefore s/he can choose any plaintexts/ciphertexts and get the corresponding ciphertexts/plaintexts. The objective of those attacks is to deduce the key [Menezes 1996].

Binary representations of variables $a$, $b$, $c$, $d$, $p$, and $q$ involved in filtering/output functions in Section 2.4 are $a_0 a_1 \ldots a_7$, $b_0 b_1 \ldots b_7$, …, $q_0 q_1 \ldots q_7$. Since $p = a \oplus c$ and $q = b \oplus d$, the following equations

$$\begin{cases} p_0 = a_0 \oplus c_0 \\ p_1 = a_1 \oplus c_1 \\ \vdots \\ p_7 = a_7 \oplus c_7 \end{cases} \text{ and } \begin{cases} q_0 = b_0 \oplus d_0 \\ q_1 = b_1 \oplus d_1 \\ \vdots \\ q_7 = b_7 \oplus d_7 \end{cases}$$

hold. In chosen-text attacks an adversary knows a keystream number $p_0 p_1 \ldots p_7 q_0 q_1 \ldots q_7$ does not mean that s/he can solve the above equations to get the state, since 16 equations are not sufficient to obtain 32 unknowns $a_0 a_1 \ldots a_7 b_0 b_1 \ldots b_7 c_0 c_1 \ldots c_7 d_0 d_1 \ldots d_7$. This S-box-like filtering function is a compression function with 200% compress ratio (32-bit-in/16-bit-out). This satisfies that the filtering function must not leak too much information on the internal state [Canteaut 2005]. Even an adversary can completely recover the all subkeys from the these attacks, s/he still have no better way than brute force to obtain the key, since reversing the key scheduling procedure is hard assuming the hash used is a good one-way function.

Often it is possible to analyze the behavior of a subsystem in dependence of a certain parameter. When key bits are directly used as parameter bits the determination of this parameter reduces the key space by the corresponding number of bits [Kelber 2005]. The key scheduling (in subsection 2.2) procedure turns the secret key into 8 (or 16) subkeys in a very complicated manner, so we avoid the problem from happening.

In distinguishing attack to stream ciphers an adversary can distinguish between the keystream of a particular cipher and the output of a truly random number generator with a non-negligible probability. In practice distinguishing attack is not a security issue [Rose 2002], therefore we do not consider it as a threat to the cipher.

## 4  Randomness Evaluation and Comparison with RC4

For randomness of keystreams produced by the Mmohocc cipher we performed two stringent batteries of statistical tests: the NIST Suite [Rukhin 2001, NIST 2001] and the DIEHARD Suite [Marsaglia 1995]. The keystreams successfully passed the randomness statistical tests with satisfactory results [Zhang 2006], see Table 3 and 4 in Appendix B.

Comparing with stream cipher RC4, Table 1 shows that the Mmohocc cipher has similar encryption speed. However the Mmohocc is using more complicated algorithm than RC4 and without key scheduling algorithm weaknesses found in RC4 [Fluhrer 2001].

Table 1   Speed Comparison with RC4

|         | 584KB  | 11,200KB | 22,400KB | 145,600KB |
|---------|--------|----------|----------|-----------|
| RC4     | 0.047s | 1.093s   | 2.109s   | 11.032s   |
| Mmohocc | 0.093s | 1.719s   | 3.468s   | 22.719s   |

## 5   Conclusion

Based on fundamental chaos characteristics of mixing, unpredictability, and sensitivity to initial conditions and multi-map orbit-hopping technique a new chaotic cipher Mmohocc is built. In the paper we discussed the major components of the cipher: chaotic map bank, key-scheduling, IV, hopping mechanism, random number extraction, and pseudo-code implementation. From security analysis in Section 3, we know that the cipher has extremely long period, $\delta$-like auto-correlations, close-to-zero cross-correlations, and it is a stochastic process of mixture of multiple Markov processes. The cipher is resistant to related-key-IV attacks, Time/Memory/Data tradeoff attacks, algebraic attacks, and chosen-text attacks. In the respect of randomness evaluation the keystreams successfully passed two most popular batteries of statistical tests and the encryption speed is similar to RC4's.

## Acknowledgements

The authors would like to thank Dr. Lin Leung for her word-by-word review to the manuscript and Professor Michael Anshel for his helpful comments and encouragement.

# Appendix A  Hopping Pattern Lookup Table

Table 2  Partial Hopping Patterns for 11 Orbits (HP-Hopping Pattern)

| HP | Orbit Permutation Sequence (from Sequential Set) | HP | Orbit Permutation Sequence (from Swapped Set) |
|---|---|---|---|
| 0 | (1,2), (3,4), (5,6), (7,8), (9,10,11) | 120 | (2,1), (4,3), (6,5), (8,7), (9,11,10) |
| 1 | (1,2), (3,4), (5,6), (9,10,11), (7,8) | 121 | (2,1), (4,3), (6,5), (9,11,10), (8,7) |
| 2 | (1,2), (3,4), (7,8), (5,6), (9,10,11) | 122 | (2,1), (4,3), (8,7), (6,5), (9,11,10) |
| 3 | (1,2), (3,4), (7,8), (9,10,11), (5,6) | 123 | (2,1), (4,3), (8,7), (9,11,10), (6,5) |
| 4 | (1,2), (3,4), (9,10,11), (5,6), (7,8) | 124 | (2,1), (4,3), (9,11,10), (6,5), (8,7) |
| 5 | (1,2), (3,4), (9,10,11), (7,8), (5,6) | 125 | (2,1), (4,3), (9,11,10), (8,7), (6,5) |
| 6 | (1,2), (5,6), (3,4), (7,8), (9,10,11) | 126 | (2,1), (6,5), (4,3), (8,7), (9,11,10) |
| 7 | (1,2), (5,6), (3,4), (9,10,11), (7,8) | 127 | (2,1), (6,5), (4,3), (9,11,10) (8,7) |
| 8 | (1,2), (5,6), (7,8), (3,4), (9,10,11) | 128 | (2,1), (6,5), (8,7), (4,3), (9,11,10) |
| 9 | (1,2), (5,6), (7,8), (9,10,11), (3,4) | 129 | (2,1), (6,5), (8,7), (9,11,10), (4,3) |
| 10 | (1,2), (5,6), (9,10,11), (3,4), (7,8) | 130 | (2,1), (6,5), (9,11,10), (4,3), (8,7) |
| 11 | (1,2), (5,6), (9,10,11), (7,8), (3,4) | 131 | (2,1), (6,5), (9,11,10), (8,7), (4,3) |
| 12 | (1,2), (7,8), (3,4), (5,6), (9,10,11) | 132 | (2,1), (8,7), (4,3), (6,5), (9,11,10) |
| 13 | (1,2), (7,8), (3,4), (9,10,11), (5,6) | 133 | (2,1), (8,7), (4,3), (9,11,10), (6,5) |
| 14 | (1,2), (7,8), (5,6), (3,4), (9,10,11) | 134 | (2,1), (8,7), (6,5), (4,3), (9,11,10) |
| 15 | (1,2), (7,8), (5,6), (9,10,11), (3,4) | 135 | (2,1), (8,7), (6,5), (9,11,10), (4,3) |
| 16 | (1,2), (7,8), (9,10,11), (3,4), (5,6) | 136 | (2,1), (8,7), (9,11,10), (4,3), (6,5) |
| 17 | (1,2), (7,8), (9,10,11), (5,6), (3,4) | 137 | (2,1), (8,7), (9,11,10), (6,5), (4,3) |
| 18 | (1,2), (9,10,11), (3,4), (5,6), (7,8) | 138 | (2,1), (9,11,10), (4,3), (6,5), (8,7) |
| 19 | (1,2), (9,10,11), (3,4), (7,8), (5,6) | 139 | (2,1), (9,11,10), (4,3), (8,7), (6,5) |
| 20 | (1,2), (9,10,11), (5,6), (3,4), (7,8) | 140 | (2,1), (9,11,10), (6,5), (4,3), (8,7) |
| 21 | (1,2), (9,10,11), (5,6), (7,8), (3,4) | 141 | (2,1), (9,11,10), (6,5), (8,7), (4,3) |
| 22 | (1,2), (9,10,11), (7,8), (3,4), (5,6) | 142 | (2,1), (9,11,10), (8,7), (4,3), (6,5) |
| 23 | (1,2), (9,10,11), (7,8), (5,6), (3,4) | 143 | (2,1), (9,11,10), (8,7), (6,5), (4,3) |
| 24 | (3,4), (1,2), (5,6), (7,8), (9,10,11) | 144 | (4,3), (2,1), (6,5), (8,7), (9,11,10) |
| 25 | (3,4), (1,2), (5,6), (9,10,11), (7,8) | 145 | (4,3), (2,1), (6,5), (9,11,10), (8,7) |

Note: the numbers within each subgroup can be arbitrarily re-arranged. For example in pattern 141, the numbers in subgroup (9, 11, 10) could be rearranged as (10, 9, 11) or as (11, 9, 10).

## Appendix B  Randomness Test Results

Table 3  Mean and Variance of *p*-values for A Large Bit Sequence for NIST Suite

| TSN | Test Name | Mean of p-value | Variance | Conclusion |
| --- | --- | --- | --- | --- |
| 1 | Approximate Entropy | 0.5146 | 0.0857 | Success |
| 2 | Block Frequency | 0.5040 | 0.0844 | Success |
| 3 | Cumulative Sums (Forward) | 0.4777 | 0.0799 | Success |
|   | Cumulative Sums (Reverse) | 0.4824 | 0.0815 | Success |
| 4 | Fast Fourier Transform (Spectral) | 0.4799 | 0.0800 | Success |
| 5 | Frequency (Mono-bit) | 0.4894 | 0.0827 | Success |
| 6 | Lempel-Ziv Compression | 0.5024 | 0.0823 | Success |
| 7 | Linear Complexity | 0.5024 | 0.0823 | Success |
| 8 | Longest Runs of Ones | 0.5121 | 0.0849 | Success |
| 9 | Maurer's Universal Statistical | 0.5000 | 0.0889 | Success |
| 10 | Non-Overlapping Template Matching | 0.4997 | 0.0833 | Success |
| 11 | Overlapping Template Matching | 0.4970 | 0.0830 | Success |
| 12 | Random Excursions | 0.5087 | 0.0835 | Success |
| 13 | Random Excursions Variant | 0.5103 | 0.0813 | Success |
| 14 | Rank | 0.4963 | 0.0841 | Success |
| 15 | Runs | 0.4980 | 0.0841 | Success |
| 16 | Serial | 0.5024 | 0.0847 | Success |

(TSN: Test Serial Number)

Table 4  *p*-values from the Diehard Statistical Test Suite

| TSN | Test Name | *p*-value | Conclusion |
| --- | --- | --- | --- |
| 1 | Birthday Spacing | 0.3213 | Success |
| 2 | Overlapping 5-Permutation | 0.1489 | Success |
| 3 | Binary Rank (31 x 31 Matrices) | 0.8363 | Success |
| 4 | Binary Rank (32 x 32 Matrices) | 0.3227 | Success |
| 5 | Binary Rank (6 x 8 Matrices) | 0.4506 | Success |
| 6 | Bitstream | 0.5638 | Success |
| 7 | Overlapping-Pairs-Sparse-Occupancy | 0.5704 | Success |
| 8 | Overlapping-Quadruples-Sparse-Occupancy | 0.5193 | Success |
| 9 | DNA | 0.4021 | Success |
| 10 | Count-The-1's (on stream of bytes) | 0.2279 | Success |
| 11 | Count0-The-1's (on specific bytes) | 0.5343 | Success |
| 12 | Parking Lot | 0.9362 | Success |
| 13 | Minimum Distance | 0.2327 | Success |
| 14 | 3D Spheres | 0.3366 | Success |
| 15 | Squeeze | 0.4369 | Success |
| 16 | Overlapping Sums | 0.6415 | Success |
| 17 | Runs | 0.5000 | Success |
| 18 | Craps | 0.5579 | Success |